\documentclass[aps,prl,twocolumn,showpacs]{revtex4}

\usepackage{graphicx}

\begin{document}


\title{Residual Energies after Slow Quantum Annealing}


\author{Sei Suzuki$^{1,2}$}
\email[]{sei@mns.k.u-tokyo.ac.jp}
\author{Masato Okada$^{1,2}$}
\affiliation{
$^1$Graduate School of Frontier Sciences, The University of Tokyo, 
Kashiwa 277-8561 \\
$^2$Brain Science Institute, RIKEN, Wako 351-0198
}


\date{\today}

\begin{abstract}
Features of the residual energy after the quantum annealing are 
investigated.
The quantum annealing method exploits quantum fluctuations
to search the ground state of classical disordered Hamiltonian.
If the quantum fluctuation
is reduced sufficiently slowly and linearly by the time,
the residual energy after the quantum 
annealing falls as the inverse square of the annealing time. 
We show this feature of
the residual energy by numerical calculations for small-sized
systems and derive it on the basis of the quantum adiabatic 
theorem.
\end{abstract}


\maketitle

Combinatorial optimization problems have presented a 
variety of challenges in vast area of sciences. 
Since the number of combination usually explodes exponentially 
with the number of elements,
they becomes unsolvable for large elements
by means of checking all of the combinations.
In particular, no algorithm suitable for conventional computers
has been discovered to solve the 
nondeterministic polynomial-time (NP) problem in a time of 
the polynomial order of elements.
The quantum annealing method has been proposed for such an
intractable problem\cite{bib:Kadowaki}.

The combinatorial optimization problem is often represented
in terms of a classical Hamiltonian, such as a random
Ising model. What we do for the problem is 
to find the ground state of the classical Hamiltonian.
In the quantum annealing method, a quantum Hamiltonian is
introduced for a classical problem.
Namely quantum fluctuations are exploited to search the solution.
The principle of the quantum annealing method 
provides us a novel algorithm for quantum computation
\cite{bib:Farhi,bib:Hogg}. In fact, fundamental experiments of quantum
computation have been carried out\cite{bib:Steffen}.
Another importance of this method lies in implementation
by conventional classical computers. 
The thermal (simulated) annealing is an well-known
method for classical computers\cite{bib:Geman}.
It has been shown experimentally that the quantum annealing
surpasses the thermal annealing\cite{bib:brookeS}.
Hence the algorithm of quantum annealing in classical computers
is expected to reduce the complexity of the problem.
However it has been in question 
whether the exponential scaling law of the complexity
regarding the problem size 
is replaced by the power law in the quantum algorithm.
It is a great issue whether this algorithm can transform
NP problems to polynomial-time solvable problems.

In this letter, we investigate the asymptotic behavior of the
residual energy for long annealing time. The residual energy
is defined by the energy difference between the 
solution by means of an algorithm and the true one.
If the scaling law of the asymptotic behavior with respect to
the process time is unveiled, it becomes clear how efficiently 
the method leads us to the true solution.
For instance, the residual energy after the thermal 
annealing decreases with the annealing (process) time as
$\Delta E \sim 1/(\ln\tau)^{\zeta}$ with $1\le\zeta \le
2$\cite{bib:Huse,bib:Shinomoto}, 
and thus extraordinarily long time is needed to obtain a closely
approximate solution.
The logarithmic scaling law of the residual energy
reminds us of the exponential law of the complexity.
For the quantum annealing method, an earlier work has estimated 
that the residual energy falls as $\Delta E\sim 1/(\ln\tau)^{\zeta}$
with $\zeta\sim 6$\cite{bib:Santoro}.
However no one has confirmed numerically this scaling law of
the residual energy after the quantum annealing.

We examine the scaling law numerically in the present work.
The result shows that the residual energy obeys power law,
$\Delta E\sim 1/\tau^{\zeta}$ with $\zeta \sim 2$ in the
long $\tau$ limit, 
in contrast to the earlier prediction.
This asymptotic behavior is derived from the adiabatic theorem of
the quantum mechanics.


Let us consider a time-dependent Hamiltonian
$\mathcal{H}(t)$
which consists of the classical Hamiltonian 
$\mathcal{H}_0$ and the tunneling Hamiltonian $\mathcal{H}_T$.
The classical Hamiltonian denotes the energy function
representing a combinatorial optimization problem,
while the tunneling Hamiltonian is responsible for 
quantum effects.
Detailed forms of these Hamiltonians are written later.
We assume the time dependence of the Hamiltonian as follows.
The tunneling term dominates the Hamiltonian at the initial time,
and it vanishes at the final time. 
The Hamiltonian changes
continuously in between the initial and final times.
There is no symmetry at any time which brings about degenerate
ground states of the Hamiltonian.
The quantum state vector stands for a solution of the combinatorial
optimization problem in the quantum annealing method.
The evolution of the state vector denoted by $|\Psi(t)\rangle$ 
with the time $t$
is determined by the Schr\"{o}dinger equation,
\begin{equation}
 i\frac{d}{dt} |\Psi(t)\rangle = \mathcal{H}(t) |\Psi(t)\rangle .
  \label{eq:Schrodinger}
\end{equation}
We assume that the initial state is adjusted to the ground state
of the initial Hamiltonian.
It follows that the approximate solution is obtained as the state vector
after an evolution with the time when the Hamiltonian coincides with the
classical Hamiltonian.
This is the process of the quantum annealing method.

The validity of the quantum annealing method originates in the
adiabatic evolution of quantum mechanical states.
The state evolving from the eigen state of the Hamiltonian
almost retains being the eigen state.
This dynamical process goes on when
the change in the Hamiltonian with the time is sufficiently slow
and the associated state is non-degenerate at any time.
Hence the state starting from the ground state of the 
initial Hamiltonian arrives at
an approximate ground state of the Hamiltonian at the final time.
In the quantum annealing method, the final Hamiltonian is
set to the classical Hamiltonian. Thus the state vector 
at the final time becomes an approximate ground state of the
classical Hamiltonian, namely the solution of the
combinatorial optimization problem.

Although we have imposed assumptions on the time dependence of
the Hamiltonian, there are some degrees of freedom on that.
We settle it as follows.
First we denote an annealing time by $\tau$
and consider the time period $[0, \tau]$.
Taking into account the assumptions in the above,
it is favored that the Hamiltonian at the initial time is identical to
the tunneling Hamiltonian. This is because it is usually easy to
make the exact ground eigen state of the tunneling Hamiltonian 
as will be seen later.
Moreover it is natural to let the Hamiltonian 
depend on the time linearly, 
since the adiabatic theorem is proven for a
Hamiltonian linear in the time. There is no strict insurance that
the adiabatic theorem works correctly for the 
Hamiltonian with other time dependences.
Thus we employ the following form of the Hamiltonian,
\begin{equation}
 \mathcal{H}_{\tau}(t) = \frac{t}{\tau}\mathcal{H}_0 + 
  \left(1 - \frac{t}{\tau}\right) \mathcal{H}_T.
  \label{eq:Htotal}
\end{equation}
Since the derivative of this Hamiltonian by the time is proportional to
$1/\tau$, the change of Hamiltonian involved by the time advance
is infinitely slow in the infinite limit of $\tau$. 
Using this Hamiltonian, the adiabatic theorem ensures
that the final state converges to the ground
eigen state of $\mathcal{H}_0$ in infinite limit of $\tau$.

We note here that the Hamiltonian (\ref{eq:Htotal}) is artificial.
Although we are interested in the classical Hamiltonian $\mathcal{H}_0$,
the Hamiltonian does not include the contribution from $\mathcal{H}_0$
initially. We emphasize here that the important thing is not
how the intermediate state is but how the final state is after
the evolution. Hence it does not matter that the Hamiltonian
at the initial time differs entirely from $\mathcal{H}_0$.


For the sake of implementation of 
the quantum annealing, we consider two simple models as pilot cases.
The first is a single-particle problem of the tight-binding model.
Let us consider a one-dimensional lattice. We assume that 
there is one energy eigen-state at each site and the states
in different sites are orthogonal each other.
The classical Hamiltonian is given by
\begin{equation}
 \mathcal{H}_0 = \sum_{i=1}^N V_i |i\rangle\langle i| ,
  \label{eq:TB1}
\end{equation}
where $|i\rangle$ denotes the single-particle state of the site $i$.
$N$ denotes the number of sites.
$V_i$ indicates the energy of the state at the site $i$. We assume
$V_i$s are randomly distributed between $0$ and $1$.
The tunneling Hamiltonian of the present system is brought by
hopping between nearest neighbor sites. It is written as
\begin{equation}
 \mathcal{H}_T = - \alpha \sum_{i=1}^{N-1} 
 \left(|i\rangle\langle i+1| + |i+1\rangle\langle i| \right) .
  \label{eq:TB2}
\end{equation}
A positive parameter $\alpha$ denotes the strength of tunneling.
The ground state of $\mathcal{H}_T$ as the initial state is
given by 
$|\Psi_0\rangle = \frac{1}{\sqrt{N}}\sum_{i=1}^N |i\rangle$.

\begin{figure}[tbp]
 \begin{center}
  \includegraphics[width=8cm,clip]{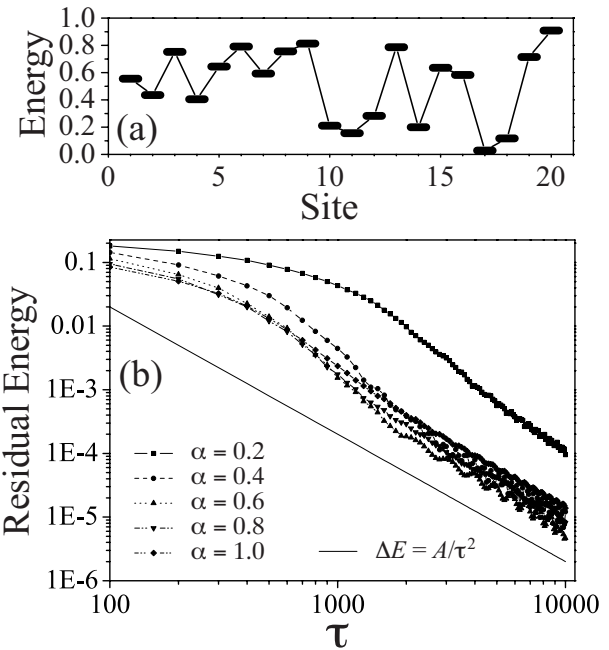}
 \end{center}
 \caption{Results on the tight-binding model. 
 Calculations were performed for the system with twenty sites, $N=20$. 
 Figure (a) shows the potential energy of each site.
 Figure (b) shows the annealing time dependence of residual energies. 
 Results for several tunneling energies are plotted. 
 A straight line, $\Delta E\propto 1/\tau^2$, is shown for
 comparison.
 Note that both axes are logarithmic.
 For large $\tau$, the residual energy obeys the power law for
 the annealing time. The exponent is estimated to be approximately two.
}
 \label{fig:TB}
\end{figure}

The next model is the random Ising spin model. The classical
Hamiltonian is written as
\begin{equation}
 \mathcal{H}_0 = - \sum_{<i,j>} J_{ij} S_i^z S_j^z - h \sum_i S_i^z ,
  \label{eq:SG1}
\end{equation}
where $S_i^z$ is the $z$ component of the spin operator at the site $i$.
The coupling constant $J_{ij}$ takes the value $+1$ or $-1$ randomly.
The Zeemann term with $h=0.1$ is included so as to avoid that 
the ground state becomes doubly degenerated.
The tunneling Hamiltonian is introduced by the transverse field.
Supposing $\alpha$ to be the transverse field, the Zeemann energy
yields the tunneling Hamiltonian,
\begin{equation}
 \mathcal{H}_T = -\alpha \sum_{i} S_i^x ,
  \label{eq:SG2}
\end{equation}
where $S_i^x$ is the $x$ component of the spin operator at the site $i$.
All spins are fully polarized along the $x$ axis in the ground state
of $\mathcal{H}_T$. It follows from this that the initial state 
is written as
$|\Psi_0\rangle = \prod_i[\frac{1}{\sqrt{2}}
\left(|\uparrow\rangle_i + |\downarrow\rangle_i\right)]$, where
$|\uparrow\rangle_i$ and $|\downarrow\rangle_i$
are the eigen states of $S_i^z$ with the eigen values $\frac{1}{2}$
and $-\frac{1}{2}$ respectively.

\begin{figure}[tbp]
 \begin{center}
  \includegraphics[width=8cm,clip]{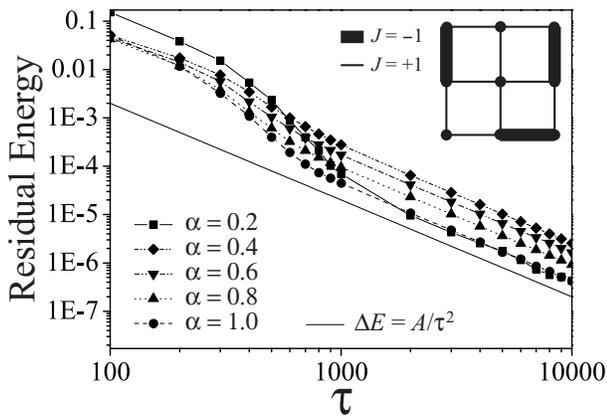}
 \end{center}
 \caption{Annealing time dependence of residual energies 
 for the random Ising spin model. Two-dimensional lattice with nine
 spins are considered. 
 Results are shown for several values of the transverse field.
 A line representing $\Delta E\propto 1/\tau^2$ is drawn for
 comparison.
 It is clear that the residual energy is fitted well by 
 the line of $\Delta E\propto 1/\tau^2$.
 The inset depicts bonds of the interaction
 and values of the coupling constant.
 }
 \label{fig:SG}
\end{figure}

For each model we solved the Schr\"{o}dinger equation 
(\ref{eq:Schrodinger}) numerically 
using the Runge-Kutta algorithm of fourth order and
obtained the time dependence of the state vector.
The residual energy is defined by the energy difference
between the energy expectation value with respect to the
final state and the ground eigen energy of
the classical Hamiltonian,
\begin{equation}
 \Delta E = \langle\Psi (\tau)|\mathcal{H}_0|\Psi(\tau)\rangle - 
 E_0,
 \label{eq:DeltaE}
\end{equation}
where $E_0$ denotes the ground eigen energy of $H_0$.
In our calculation, $E_0$ is known because the system
is small enough.

We show results on the tight-binding model represented by
eqs. (\ref{eq:TB1}) and (\ref{eq:TB2}) in Fig.\ref{fig:TB}.
The potential 
energy at each site is randomly generated and is shown in 
Fig\ref{fig:TB}(a). Calculations ware performed for 
several tunneling energies. Large value of $\alpha$
gives rise to the large energy gap at the initial time.
Since the adiabatic theorem works better for larger energy gap,
large $\alpha$ is preferable to lower the residual energy.
However large $\alpha$ yields large derivative by the time of the
Hamiltonian. Since large derivative disturbs
the adiabatic evolution, large $\alpha$ is not preferable.
Therefore there is an optimum $\alpha$.
In the present case, it seems around $\alpha = 0.6$.

It is seen in Fig.\ref{fig:TB}(b) that
the logarithm of residual energy decreases almost linearly with 
increasing the logarithm of annealing time for long annealing time. 
Therefore the residual energy roughly obeys 
the power law with respect to the annealing time, 
\begin{equation}
 \Delta E \propto \frac{1}{\tau^{\zeta}}.
\label{eq:SquareInv}
\end{equation}
By comparison with the line 
$\Delta E \propto\tau^{-2}$ drawn in the same figure,
the exponent is estimated as $\zeta\sim 2$.

Results on the random Ising model given by eqs.(\ref{eq:SG1}) and 
(\ref{eq:SG2}) are shown in Fig.\ref{fig:SG}. The system consists of
nine spins in the two-dimensional square lattice with nearest neighbor
interactions. The values of coupling constants are indicated by the
inset in the figure. 
Similar to the case in Fig.\ref{fig:TB}(b), the residual energy
is smallest at a certain optimum $\alpha$ for a fixed $\tau$. 
It is clearly seen that the logarithm of residual energy is
linear in $\ln\tau$ for long $\tau$.
Hence the power law of the residual energy, eq.(\ref{eq:SquareInv})
\begin{equation}
 \Delta E \sim \frac{1}{\tau^{\zeta}},
\end{equation}
holds also in the present case. 
Furthermore we can estimate $\zeta\sim 2$ by comparison with 
the straight line
of $\Delta E\propto 1/\tau^2$ in the figure.
This result is the same as that of the tight-binding model.

From results of numerical calculations, we naturally infer
that the power law with the exponent $\zeta\sim 2$ is 
the universal feature in the quantum annealing method.
In fact, one can derive from the adiabatic theorem that this is true.
According to a rigorous theory on the adiabatic 
theorem,
the probability to find out the excited states of $\mathcal{H}_0$ in
the final state is given by\cite{bib:Messiah}
\begin{equation}
 \sum_{n\neq 0} |\langle n|\Psi(\tau)\rangle|^2 
  \sim O\left(\frac{1}{\left(\epsilon\tau\right)^2}\right)
  \label{eq:error}
\end{equation}
where $|n\rangle$ denotes the eigen 
state of $\mathcal{H}_0$ and $n=0$ indicates the ground state
in particular. 
$\epsilon$ has the dimension of energy and defined by
$\epsilon = \varepsilon_{\rm min}^2/\varepsilon'$
where $\varepsilon_{\rm min}$ is the minimum of energy gap
above the instantaneous ground state of $\mathcal{H}_{\tau}(t)$
and $\varepsilon'$ is the maximum of matrix element of 
$\mathcal{H}_T - \mathcal{H}_0$ among instantaneous ground state
and the lowest excited state of $\mathcal{H}_{\tau}(t)$.
The value of $\varepsilon'$ is of the order of the energy unit.
Equation (\ref{eq:error}) holds for $\epsilon\tau\gg 1$. 
Formulas for each eigen state are derived, that is, 
$|\langle n|\Psi(\tau)\rangle|^2 
\sim O(1/\left(\epsilon\tau\right)^2)$ for
$n\neq 0$ and $|\langle 0|\Psi(\tau)\rangle|^2 
\sim 1 - O(1/\left(\epsilon\tau\right)^2)$.
Hence, arranging the eq.(\ref{eq:DeltaE}), 
we obtain that the residual energy 
is of the order of $1/\left(\epsilon\tau\right)^2$.
\begin{eqnarray}
 \Delta E &=& \sum_{n\neq 0} E_n |\langle n|\Psi(\tau)\rangle|^2 - 
  E_0 \left(1- |\langle 0|\Psi(\tau)\rangle|^2\right) \nonumber \\
  &\sim& O(\frac{1}{\left(\epsilon\tau\right)^2}) ,
  \label{eq:DeltaE1}
\label{eq:powerlaw}
\end{eqnarray}
where $E_n$ is the eigen energy of $\mathcal{H}_0$.
Thus it is shown that the term proportional to $1/\tau^2$ dominates the
residual energy in the infinite limit of $\tau$.

We note that eq.(\ref{eq:powerlaw}) may contain higher order terms
of $1/\left(\epsilon\tau\right)$ than $1/\left(\epsilon\tau\right)^2$.
We attribute fluctuations around the line of
$\Delta E\propto 1/\tau^2$ in Fig.\ref{fig:TB}(b) to these 
additional terms.

The existence of the energy gap above the instantaneous ground
state of $\mathcal{H}_{\tau}(t)$ is crucial
for the quantum annealing method. 
The argument on the basis of the eq.(\ref{eq:error}) is
incorrect when the ground state is degenerate at
a time during the evolution.
This is obvious because of $\epsilon=0$.
However the Hamiltonian of disordered systems 
does not usually possess any symmetry 
which brings about degenerate ground state.

We assume that the instantaneous energy gap is always finite.
The value of $\varepsilon_{\rm min}$ depends on the problem. 
For problems with small $\varepsilon_{\rm min}$, 
a large $\tau$ is required in order to validate the power 
law of the residual energy.
In other words, 
the quantum annealing method is more efficient for problems
with larger $\varepsilon_{\rm min}$.

For $\epsilon\tau\sim 1$, 
the adiabatic theorem does not work.
Instead the Landau-Zener theory should account for 
the transition between two states with close energies.
Consider that the energy levels of instantaneous ground and 
first excited states come close together at a time.
The excitation energy is locally minimized at this time.
According to the Landau-Zener theory\cite{bib:LandauQM,bib:Zener}, 
the transition probability
from the ground state to the first excited state is given
by $p\sim \exp(-c \epsilon\tau)$, 
where $c$ is a constant of the order of unity.
If such a possible transition occurs just once,
the residual energy is proportional to $\exp(-c \epsilon\tau)$.
However it is not obvious whether this exponential behavior
is true for large systems, since the transition events
may take place many times.

Santoro \textit{et al.} have predicted the logarithmic scaling
law of the residual energy
on the basis of the Landau-Zener theory\cite{bib:Santoro}.
We remark that our result of power law is valid for 
$\tau\gg 1/\epsilon$ where the quantum adiabatic theorem
is guaranteed. Since the Landau-Zener theory is intended for
$\tau\sim 1/\epsilon$, these two scaling laws may not conflict
with each other.
Nevertheless it should be noted that the definition of the
residual energy in ref.\cite{bib:Santoro} is given 
by the expectation value of the excitation energy
among the quantum annealing process. 
It is natural that the residual energy
is defined by the energy expectation value of the final state 
measured from the true ground energy, eq.(\ref{eq:DeltaE}).
Furthermore no one has confirmed numerically the scaling behavior 
for $\tau\sim 1/\epsilon$.
In order to unveil the feature in this regime,
numerical calculations for huge sized systems are desired.

We remark on the implementation of the quantum annealing
for combinatorial optimization problems with large elements.
In order to carry out the quantum annealing, we must 
compute the real-time dynamics according to the Schr\"{o}dinger
equation. Methods of numerical calculation for the real-time
quantum dynamics have not 
been developed enough except for cases of one-dimensional models. 
Construction of a general method is a big issue.
One possibility exists in the quantum Monte-Carlo
method. Even though direct computations of the real-time
dynamics are impossible in the quantum Monte-Carlo method,
one can regard an evolution by Monte-Carlo step 
as a fictitious-time evolution.
some of earlier works have 
reported that
the quantum annealing with the quantum Monte-Carlo
is more efficient than the 
thermal annealing\cite{bib:Kadowaki,bib:Martonak}.
However it is not clear that the adiabatic theorem works 
in the Monte-Carlo evolution, 
the residual energy may show annealing time dependence different from
real-time quantum annealing.
A rigorous theory on the residual energy after the 
Monte-Carlo evolution is left as a future work.



The present work is supported by Grant-in-Aid 
for Scientific Research (Grant No.14084212) 
by the Ministry of Education, Culture, Sports, Science and 
Technology of Japan.

\bibliography{letter}

\end{document}